\documentclass{elsarticle}

\usepackage{amsmath,amsfonts,amsthm}
\usepackage{graphicx}
\usepackage{multirow}
\usepackage{tikz}
\usepackage{soul}
\usepackage{url}
\usetikzlibrary{arrows.meta}
\usetikzlibrary{angles,quotes}
\newcommand{\beq}{\begin{equation}}
\newcommand{\eeq}{\end{equation}}

\DeclareMathOperator*{\argmin}{arg\,min}

\newtheorem{prop}{Proposition}

\title{Two Dimensional Magnetic Current Imaging Via $L1-\textrm{Curl}$ Regularized Divergence Free Wavelet Reconstruction}
\author{Christopher Miller\corref{miller}\fnref{mitre}}
\author{Jacob Lenz\fnref{mitre}}
\author{Adrian Mariano\fnref{mitre}}
\author{Dmitro Martynowych\fnref{mitre,mit}}
\author{Sean Oliver\corref{oliver}\fnref{mitre}}

\cortext[miller]{cwmiller@mitre.org}
\cortext[oliver]{smoliver@mitre.org}
\fntext[mitre]{The MITRE Corporation}
\fntext[mit]{The Massachusetts Institute of Technology, Laboratory for Information and Decision Systems}

\begin{document}

\bibliographystyle{elsarticle-num}

\begin{abstract}

The reconstruction of current distributions from samples of their induced magnetic field is a challenging problem due to multiple factors. First, the problem of reconstructing general three dimensional current distributions is ill-posed. Second, the current-to-field operator performs a low-pass filter that dampens high-spatial frequency information, so that even in situations where the inversion is formally possible, attempting to employ the formal inverse will result in solutions with unacceptable noise. Most contemporary methods for reconstructing current distributions in two dimensions are based on Fourier techniques and apply a low pass filter to the $B$-field data, which prevents excessive noise amplification during reconstruction at the cost of admitting blurring in the reconstructed solution. In this report, we present a method of current recovery based on penalizing the $L1$ norm of the curl of the current distribution. The utility of this method is based on the observation that in microelectronics settings, the conductivity is piecewise constant. We also reconstruct the current fields using a divergence-free wavelet basis. This has the advantage of automatically enforcing current continuity and halving the number of unknowns that must be solved for. Additionally, the curl operator can be computed exactly and analytically in this wavelet expansion, which simplifies the application of the $L1-\textrm{curl}$ regularizer. We demonstrate improved reconstruction quality relative to Fourier-based techniques on both simulated and laboratory-acquired magnetic field data.
\end{abstract}


\maketitle

\section{Introduction}
Magnetic field imaging (MFI) of microelectronics is a powerful technique for fault detection and other inspection tasks due to its sensitivity and non-destructive nature. In spite of these advantages and the continual increase in magnetometer quality, current distribution reconstruction from MFI remains a difficult task as the scale of features in microelectronic devices continues to shrink. An additional emerging difficulty is the increasing use of 3D integrated circuits wherein there are multiple electronically active layers. 

It is well known that the general problem of reconstructing a $3-d$ current distribution from magnetic field measurements is ill-posed. As such, potential solution methods must make some assumptions regarding the structure of the underlying current distribution. Another difficulty is the fact that even for the $2-d$ case, the magnetic field formed by a $2-d$ current distribution is an aggressively low-passed image of that current distribution. This makes reconstruction of high -spatial frequency components of the current distribution image challenging in the presence of noise. The suppression of high-frequency content is exacerbated by increasing the standoff distance between the magnetic field sensor and the source current distribution, which makes MFI inspection impractical at large standoff distances.

In this report, we present a method of regularizing the reconstruction of current distributions that preserves high spatial frequency information present in the current distribution, is robust to both noise as well as increases in the measurement standoff distance, and relies on a very modest assumption regarding the nature of the conducting substrate. Specifically, our chosen regularizer is based on the fact that the conductors in microelectronic devices are homogeneous and, as such, the conductivity can be modeled as piecewise constant. As a consequence of Maxwell's equations, the curl of the current distribution is zero almost everywhere. In a discretized setting, this corresponds to an assumption that the curl is sparse, i.e., the majority of elements have zero curl. Regularizers that enforce a sparse gradient structure on the recovered solution have a rich history in image reconstruction in the presence of noise. In particular, use of our proposed regularizer draws heavily on the works of \cite{rudin, chambolle}, who employed a regularizer based on the total-variation semi-norm to reconstruct noisy images. The motivation for this regularizer is based on the observation that most natural images are piecewise constant and, therefore, the gradient has sparse structure. We hypothesize that penalizing the $L1$-norm of the curl represents a natural extegeneralizationsion of total-variation denoising to the domain of conservative flow fields. Efficiently computable regularizedrs least squares formulations that enforce sparsity on some feature of the solution in turn grew out of the literature on compressed sensing \cite{donoho, tao}.

In addition to assumptions regarding the homogeneity of the conducting substrate, it can also be safely assumed that, due to conservation of charge, the current density is divergence free. Thus, any computed solution should be divergence free as well. One straightforward way of enforcing this would be to add another regularization term (or optimization constraint), which would penalize solutions that had non-trivial divergence terms. There are two primary difficulties here. First, each regularizer and/or constraint adds complexity to the optimization scheme that will be employed to find the desired solution. Second, in the discrete setting, one is forced to approximate the discrete divergence using finite differences, which introduces a host of stability issues. To sidestep these issues, we restrict the space of possible solutions to be divergence free by use of divergence-free vector wavelet bases \cite{lemarie,Harouna2013,kadri}. There are a multitude of advantages to this approach. First, conservation of charge is a guarenteed property of any returned solution. Second, since the basis functions employed are vector valued, $2-d$ current distributions can be expressed using half the number of unknown parameters. Viewed a different way, the divergence-free constraint removes half of the degrees of freedom from the problem with all the attendant benefits in reduced computation time and improved noise robustness. Third, the particular construction of the wavelet bases allows for partial derivatives to be computed analytically, allowing us to sidestep the need to employ discrete difference operators. This last property is employed in computing the curl-based regularization term discussed above.

The following section begins with a full description of the $2-d$ magnetic inverse problem and our proposed approach. Following the description of the methods, we demonstrate improved reconstruction as a function of noise and measurement standoff distance relative to Fourier techniques on simulated data. We also demonstrate the technique applied to magnetic field data acquired in our microelectronics lab using a quantum diamond microscope (QDM) magnetometer \cite{Turner2020, levineturner}.

\section{Problem Formulation}\label{sec:formulation}
\subsection{Description of the Forward Model}
We begin by deriving the forward model that maps current distributions to observed magnetic fields. Assume that there is a two-dimensional current distribution, 
\begin{equation*}
    J(\textbf{x} = [x,y]^t) = [J_x, J_y],
\end{equation*}
distributed through a slab of conducting material of thickness $d$. The slab is assumed to be at height $z=0$. Assume a magnetic field measurement device performs measurements in a plane located at height $z > 0$. Following \cite{wellstood, roth}, the components of the magnetic field are given by
\begin{gather}\label{2d-Biot}
\begin{split}
B_x(\textbf{x}) &= \frac{\mu_0 z d}{4\pi}\int\int_{\mathbb{R}^2} \frac{J_{y}(\textbf{x}')}{(||\textbf{x} - \textbf{x}'||^2 + z^2)^{3/2}} \, d\textbf{x}'
    \\
    B_y(\textbf{x}) &= -\frac{\mu_0 z d}{4\pi}\int\int_{\mathbb{R}^2} \frac{J_{x}(\textbf{x}')}{(||\textbf{x} - \textbf{x}'||^2 + z^2)^{3/2}} \, d\textbf{x}'
    \\
    B_z(\textbf{x}) &= \frac{\mu_0 z d}{4\pi}\int\int_{\mathbb{R}^2} \frac{J_x(\textbf{x}')(y-y') - J_y(\textbf{x}')(x-x')}{(||\textbf{x} - \textbf{x}'||^2 + z^2)^{3/2}} \, d\textbf{x}'.
\end{split}
\end{gather}
The $x$ and $y$ components of the magnetic field are seen to be the two-dimensional convolution of the current distribution $J$ with the kernel $G$ defined by
\begin{equation}
G(\textbf{x},z) = \frac{\mu_0z d}{4\pi} \frac{1}{(||\textbf{x}||^2 + z^2)^{3/2}},
\end{equation}
which has Fourier transform 
\begin{equation}\label{ft-g}
g(\textbf{k},z) = \frac{\mu_0 d}{2} \exp(-2\pi z||\textbf{k} ||).
\end{equation}
(See Appendix \ref{app:B} for derivation).
 By the convolution theorem we obtain,
\begin{gather}\label{b-xy}
\begin{split}
b_{x}(\textbf{k},z) &= g(\textbf{k},z)j_{y}(\textbf{k}),
\\
b_{y}(\textbf{k},z) &= -g(\textbf{k},z)j_x(\textbf{k}).
\end{split}
\end{gather}
Likewise, we have that
\begin{equation}\label{b-z}
    b_{z} = i g(\textbf{k},z)\left(\frac{k_x}{||\textbf{k}||} j_y(\textbf{k}) - \frac{k_y}{||\textbf{k}||} j_x(\textbf{k})\right), 
\end{equation}
which follows from \eqref{ft-g} and \eqref{ft-gx}. Thus given $b_x$ and $b_y$, $j$ is formally computable by simply deconvolving with respect to the kernel $G$. Before concluding that the problem is trivial, one should note that the convolution kernel strongly attenuates components of the current distribution with high -spatial frequency due to the exponential roll-off present in \eqref{ft-g}. As a result of the exponential decay of high-|$\textbf{k}$ components, inversion based on simple deconvolution becomes unstable in the presence of measurement noise and large standoff distance $z$. 

To avoid excessive amplification of measurement noise, a standard recovery method is to first apply a cosine taper to the measured field data \cite{roth}. 
\begin{gather}\label{cos-taper}
\begin{split}
    \hat{j}_{x,\mathcal{F}} &= -g^{-1}\mathcal{C}\hat{b_y}
    \\
    C[k_x,k_y] &= \left[\begin{array}{cc}\frac{(1+\cos(\pi|k_x|/k_{max}))(1+\cos(\pi|k_y|/k_{max}))}{4} & |k_x|,|k_y| \leq k_{max}
    \\
    0 & \text{otherwise},
\end{array}\right.
\end{split}
\end{gather}
(similar relations hold for $b_y$ and $b_z$). The parameter $k_{max}$ defines the maximum spatial frequency for which recovery is attempted and can be considered as a type of regularization parameter. The cosine taper prevents noise amplification associated with the exponential falloff in \eqref{ft-g} at the cost of reducing the effective spatial resolution of the recovered image.

As an aside, we remark that by combining \eqref{b-z} with the continuity condition 
\begin{equation}
    i\textbf{k}\cdot [j_x(\textbf{k}),j_y(\textbf{k})] = 0,
\end{equation}
it is formally possible to compute $j_x$ and $j_y$ (modulo $\textbf{k} = 0$) with only knowledge of $b_z$ \cite{roth}. Throughout the remainder of this study, we assume that the problem is posed such that all three vector components of the magnetic field are available. This corresponds to the vector measurement capability provided by the QDM that our experimental results are based on \cite{Turner2020,levineturner}. We expect that the curl regularizarier proposed below will also have utility in the case where only a single field component is available.

Let $\textbf{J} = [\textbf{J}_x | \textbf{J}_y]^t\in \mathbb{R}^{2(N\times N)}$ be a real valued vector containing both components of the current density $J$ sampled on a $N\times N$ uniform grid with spacing $\Delta x$ in each direction. Likewise, let $\textbf{B}\in \mathbb{R}^{3(N\times N)}$ be the vector of samples of the magnetic field according to \eqref{2d-Biot}. We define the discrete forward operator $\mathcal{B}$ as
\begin{gather}\label{forward-operator}
    \begin{split}
        \textbf{B} &= \mathcal{B} \textbf{J},
        \\
        \mathcal{B} &= \left[\begin{array}{c|c|c}
\mathcal{F}^{-1} & \mathbf{0} & \mathbf{0}
\\
\hline
\mathbf{0} & \mathcal{F}^{-1} & \mathbf{0} \\
\hline
\mathbf{0} & \mathbf{0} & \mathcal{F}^{-1}
\end{array}\right]\left[\begin{array}{c|c}
\mathbf{0} & \mathcal{G}
\\
\hline
-\mathcal{G} & \mathbf{0}
\\
\hline
\mathcal{G}_x & \mathcal{G}_y
\end{array}\right] \left[\begin{array}{c|c}
\mathcal{F} & \mathbf{0}
\\
\hline
\mathbf{0} & \mathcal{F}\end{array}\right],
\\
\mathcal{G} &= diag(g[\textbf{k},z]), \mathcal{G}_x = -i\ diag\left(g[\textbf{k},z]\frac{k_y}{||\textbf{k}||}\right),\mathcal{G}_y = i\ diag\left(g[\textbf{k},z]\frac{k_x}{||\textbf{k}||}\right),
    \end{split}
\end{gather}
where $\mathcal{F}$ and $\mathcal{F}^{-1}$ are the forward and inverse $2-d$ discrete Fourier transform respectively.

\subsection{Description of the Proposed Regularization Scheme}
We assume that a device measures the $B$ field following \eqref{forward-operator} and that the resulting measurements are corrupted by additive zero-mean Gaussian noise, i.e., we have measurements $\hat{\textbf{B}}$ such that
\begin{equation}\label{noisy-measurement-model}
\hat{\textbf{B}} = \mathcal{B}\textbf{J} + \eta,\ \eta \sim N(0,\sigma^2 I).
\end{equation}
Given $\hat{\textbf{B}}$, we propose to find $\textbf{J}$ as the minimizer of a regularized least squares error functional
\begin{equation}\label{functional-1}
    \hat{\textbf{J}} = \argmin_{\textbf{J}} ||\mathcal{B}\textbf{J} - \hat{\textbf{B}}||^2 + \tau ||\textrm{curl}(\hat{\textbf{J}})||_{L_1},
\end{equation}
subject to $\hat{\textbf{J}}$ satisfying the current continuity constraint. In \eqref{functional-1}, $\tau>0$ is a regularization parameter that controls the relative importance of the regularization term. For situations where the measurement noise is small, $\tau$ should be set small so that more importance is placed on recovering a solution that fits the data. Conversely, as the noise grows larger, $\tau$ should increase to prevent recovered solutions that overfit to noisy data. The curl operator above represents a discretization of the continuous curl operator applied to the discretely sampled field $\textbf{J}$. Our approach for discretization of the curl operator will be discussed in the following subsection. 

The motivation for the curl term is based on the fact that within a micro-electronic device, it can be safely assumed that the conducting pathways are homogeneous conductors, i.e., the conductivity $\sigma_J$ is a piecewise constant function. From Maxwell's equations we have that
\begin{equation}
    \textrm{curl}(\textbf{J}) = \textrm{curl}(\sigma_J E) = -\textrm{curl}(\sigma_J \nabla V) =_{a.e.} -\sigma_J \textrm{curl}(\nabla V) = 0,
\end{equation}
where the third equality holds almost everywhere in the sense of Lebesgue measure. In a discrete setting, $\textrm{curl}(\textbf{J}) = 0$ almost everywhere implies that discrete evaluations of the curl will be sparse, having non-zero value only where there are discontinuities in the conductivity (e.g., the boundary of conductive traces). It is well known that $L1$ penalty functions tend to enforce sparsity \cite{tao,donoho}. In particular, we consider the regularizer proposed above as an analog to the total-variation regularizers that have been successfully employed to image reconstruction tasks \cite{chambolle,rudin}. Here, instead of making the assumption that an image is piecewise constant and then penalizing the $L1$ norm of the image gradients, we are assuming piecewise constant diffusivity in a conservative flow and penalizing the $l1$ norm of the curl. Regularizers based on curl penalization have also been used in other flow approximation/reconstruction tasks, both in fluid dynamics and image processing \cite{Heitz2010VariationalFF}.

\subsection{Divergence-Free Wavelet Bases}

In the formulation above, we constrain the reconstructed field $\hat{\textbf{J}}$ to satisfy the current continuity condition. We enforce this by expressing $\textbf{J}$ using a divergence-free wavelet expansion; that is, $\textbf{J}$ is expressed in a wavelet basis that contains only functions that are divergence free. A full discussion of wavelet theory is outside the scope of this report, but for a complete description of the construction of divergence free wavelets, see \cite{lemarie,Harouna2013,kadri}. For a more general discussion of wavelet theory see \cite{mallat2008wavelet}. We will present an abbreviated discussion here for completeness.

A biorthogonal multiresolution analysis can be defined in terms of four functions, the primal wavelet and scaling functions $(\psi,\phi)$ and the dual wavelet and scaling functions $(\tilde \psi, \tilde \phi)$. Define
\begin{gather}
    \begin{split}
        \phi_{j,k} = 2^{j/2}\phi(2^jx - k)
        \\
        \psi_{j,k} = 2^{j/2}\psi(2^jx-k),
    \end{split}
\end{gather}
to be the family of functions defined by scaling and translating the wavelet and scaling function. The biorthogonality condition implies that
\begin{gather}
    \begin{split}
        \langle \phi_{j,k},\tilde{\phi}_{j,k'}\rangle &=  \delta_{k,k'}
        \\
        \langle \psi_{j,k},\tilde{\psi}_{j',k'}\rangle &= \delta_{k,k'}\delta_{j,j'}
        \\
        \langle \phi_{j,k},\tilde{\psi}_{j,k}\rangle &= 0
        \\
        \langle \tilde{\phi}_{j,k},\psi_{j,k}\rangle &= 0,
    \end{split}
\end{gather}
where $\langle \cdot,\cdot\rangle$ denotes the $L^2$ inner product. The fundamental result that enables construction of divergence-free vector-valued wavelet bases is the following proposition from \cite{lemarie} that we restate as written in \cite{kadri};
\begin{prop}\label{div-free-prop}
Let $(\phi^1, \tilde\phi^1)$ be a pair of biorthogonal scaling functions associated to biorthogonal wavelets $(\psi^1,\tilde\psi^1)$, with $\phi^1 \in \mathcal{C}^{1+\epsilon}(\mathbb{R})$. There exists biorthogonal primal and dual scaling functions $(\phi^0,\tilde\phi^0)$ and biorthogonal primal and dual wavelets $(\psi^0,\tilde\psi^0)$, satisfying:
\begin{gather}\label{div-free-rel}
\begin{split}
\frac{d}{dx}\phi^1(x) = \phi^0(x) - \phi^0(x-1),
\\
\frac{d}{dx}\tilde\phi^0(x) = \tilde\phi^1(x+1) - \tilde\phi^1(x),
\\
\psi^1(x) = 4\int_{-\infty}^x\psi^0,
\\
\tilde\psi^0(x) = -4\int_{-\infty}^x\tilde\psi^1.
\end{split}
\end{gather}
\end{prop}

This proposition establishes the existence of wavelet bases linked by differentiation. Furthermore, \cite{lemarie} provides an explicit formula for the construction of a set of filter banks for the related wavelet and scaling function $(\phi_0,\psi_0,\tilde\phi_0,\tilde\psi_0)$ given the filter banks associated with $(\phi_1,\psi_1,\tilde\phi_1,\tilde\psi_1)$. This allows a straightforward implementation of the fast biorthogonal wavelet transform associated with the family $(\phi_0,\psi_0,\tilde\phi_0,\tilde\psi_0)$ when given filter coefficients for the fast wavelet transform associated with $(\phi_1,\psi_1,\tilde\phi_1,\tilde\psi_1)$. 

Typically, analysis of $2-d$ images using wavelet techniques is achieved with the isotropic $2-d$ wavelet transform. In this formulation, the basis functions are parameterized by a single scale parameter $j$. Functions of two variables are then expressed as linear combinations of functions of the form
\begin{gather}
    \begin{split}
        &\phi_{j,k}(x)\phi_{j,k'}(y)
        \\
        &\phi_{j,k}(x)\psi_{j,k'}(y)
        \\
        &\psi_{j,k}(x)\phi_{j,k'}(y)
        \\
        &\psi_{j,k}(x)\psi_{j,k'}(y).
    \end{split}
\end{gather}
The construction of divergence-free wavelets is made more convenient by instead using the anisotropic (sometimes called the fully separable) wavelet transform. This formulation uses a scale parameter $\textbf{j} = j_x,j_y$ that can be different along each axis, with the benefit that the two dimensional wavelet transform can now be expressed in terms of the wavelet function $\psi$ only. This simplifies the construction of divergence free wavelets using the relations established in \eqref{div-free-rel}. 

We define the basis functions associated with the anisotropic divergence free wavelet transform by

\begin{gather}\label{divergence-free}
    \begin{split}
         \Psi_{\textbf{j,k}} &= \left| \begin{array}{c} 
    \psi_{j_1,k_1}^1 \otimes (\psi_{j_2,k_2}^1)'
    \\
    -(\psi_{j_1,k_1}^1)'\otimes \psi_{j_2,k_2}^1
    \end{array}\right.
    \\
    &= \left| \begin{array}{c} 
    2^{j_2+2}\psi_{j_1,k_1}^1 \otimes \psi_{j_2,k_2}^0
    \\
    -2^{j_1+2}\psi_{j_1,k_1}^0\otimes \psi_{j_2,k_2}^1
    \end{array}\right. ,
    \end{split}
\end{gather}
where the second equality follows from Proposition \ref{div-free-prop}. Note that the vector valued functions in \eqref{divergence-free} are divergence-free by construction. We constrain our reconstructed current fields to have the form
\begin{equation}\label{full-decomp}
    \textbf{J}(x,y) = \sum_{j_1,k}c_{x,j_1,k}\left|\begin{array}{c}\psi^0_{j_1,k}(y) \\ 0\end{array}\right. +\sum_{j_2,k}c_{y,j_2,k}\left|\begin{array}{c} 0 \\ \psi^0_{j_2,k}(x)\end{array}\right. + \sum_{\textbf{j,k}}c_{\textrm{div},\textbf{j,k}}\Psi_\textbf{j,k}.
\end{equation}
The first two terms denote those portions of the current distribution that are constant in $x$ and $y$ respectively. We denote the forward fast discrete divergence-free wavelet transform $\mathcal W$ and its inverse as $\mathcal W^{-1}$. Following \cite{Harouna2013, kadri}, the forward and backwards transforms admit an efficient implementation that requires only standard fast wavelet transforms and a change of basis.

One additional benefit associated with the use of this formulation is that for a function defined by \eqref{full-decomp}, the curl of that function can be evaluated analytically. To accomplish this we apply Proposition \ref{div-free-prop} again to the family $(\phi_0,\psi_0,\tilde \phi_0,\tilde \psi_0)$ to derive another biorthogonal wavelet family  $(\phi_{-1},\psi_{-1},\tilde \phi_{-1},\tilde \psi_{-1})$. Thus we have that
\begin{equation}
\frac{\partial\textbf{J}_y}{\partial x} = \sum_{j_2,k}2^{j_2+2}c_{y,j_2,j}\psi^{-1}_{j_2,k}(x) + \sum_{j,k}2^{j_1+2}(-2^{j_1+2})c_{div,\textbf{j},\textbf{k}}\psi^{-1}_{j_1,k_1}\otimes\psi^1_{j_2,k_2}.
\end{equation}
In a similar fashion we can analytically find any other partial derivative. Taking a derivative of a function expressed by \eqref{full-decomp} is thus equivalent to reconstructing that function in a basis related to the original through Proposition \ref{div-free-prop}. This approach is employed in \cite{kadri} to perform high-order regularization for reconstruction of simulated fluid flows.
We can now restate the objective function \eqref{functional-1} in terms of the divergence-free wavelet expansion as
\begin{gather}\label{functional-2}
\begin{split}
    \hat{\textbf{W}} = \argmin_{\textbf{W}} ||\mathcal{B} \mathcal{W}^{-1} \textbf{W} - \hat{\textbf{B}}||^2 + \tau ||\textrm{curl}(\mathcal{W}^{-1}\textbf{W})||_{L1}
    \\
    \hat{\textbf{J}} = \mathcal{W}^{-1} \hat{\textbf{W}},
    \end{split}
\end{gather}
where $\textbf{W}$ denotes the coefficients of the expansion and the curl operator is computed using the technique discussed above.

\section{Implementation Details and Experimental Results}

\subsection{Data Simulation}
\begin{figure}
    \centering
    \includegraphics[scale=0.5]{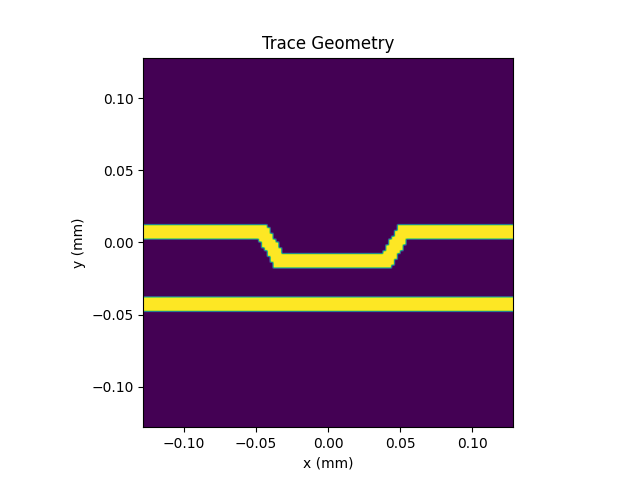}
    \caption{Geometry of traces for simulated data cases. In simulation, the top trace current is flowing left to right, while the bottom trace is flowing from right to left.}
    \label{fig:trace_geometry}
\end{figure}

In order to study the effects of varying noise levels and measurement standoff distance, we constructed a simple simulated current density and then applied the forward operator to compute simulated $B$ fields. The geometry of the simulation is defined by two conductive traces, one straight and one with a bend, running parallel to one another, as shown in Figure \ref{fig:trace_geometry}.
Traces are each 10 microns wide and they are separated by 20 microns at their closest point. We specify the conductivity of each trace to be that of copper $\sigma_J = 5.98\times 10^7 (\textrm{S/m})$. These parameters were chosen to be representative of power rail dimensions on recent generations of micro-electronic devices.

The measurement domain is defined to be a square measuring $0.256\ \textrm{mm}$ on each side. The distance between samples in the simulation is $2$ microns. Stability of the Poisson equation requires strictly positive diffusivity at each point in the domain. Therefore, for points in the spatial domain that are not part of one of the traces we set the conductivity to a very small value, $\sigma_J = 10^{-6} (\textrm{S/m})$.

\begin{figure}
    \centering
    \includegraphics[scale=0.5]{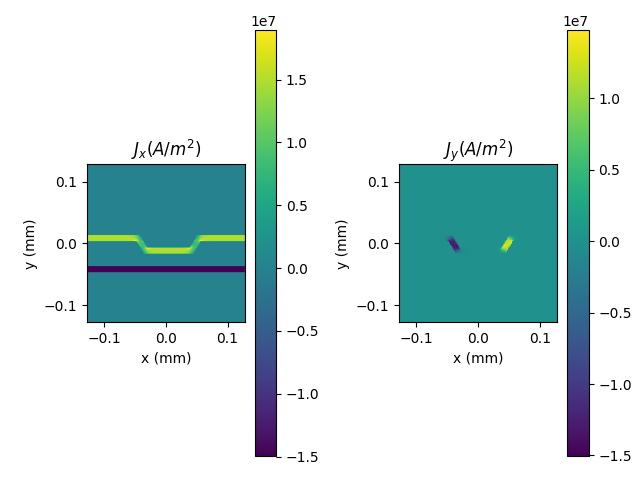}
    \caption{Simulated current density}
    \label{fig:simulated_density}
\end{figure}

In order to simulate a realistic current distribution on the traces we used a finite volume solver \cite{fipy} to solve Poisson's equation,
\begin{equation}\label{poisson}
    -\nabla\cdot(\sigma_J \nabla V) = 0,
\end{equation}
on the simulation domain. Once we solve for the voltage $V$, the current density can be computed by the relation, $\textbf{J} = -\sigma_J\nabla V$. The solution of \eqref{poisson} depends on setting appropriate boundary conditions. We set a Dirichlet boundary condition on the left hand side of the domain so that $V=0$. We set a Neumann condition on the right hand side of each trace to simulate a fixed current draw of $160 \mu \textrm{A}$ on each trace. For the purposes of modeling, the conducting layer was assumed to be $1\mu \textrm{m}$ thick. This, combined with the trace thickness and the prescribed current draw led to nominal current densities of $J_0 = 1.6\times 10^7 \textrm{A}/\textrm{m}^2$ on each trace. The current on the top trace is arranged to flow from left to right, and the bottom trace flows from right to left. All other points of the boundary have a zero-flux boundary condition applied. The simulated current density is shown in Figure \ref{fig:simulated_density}. 

\begin{figure}
    \centering
    \includegraphics[scale=0.755]{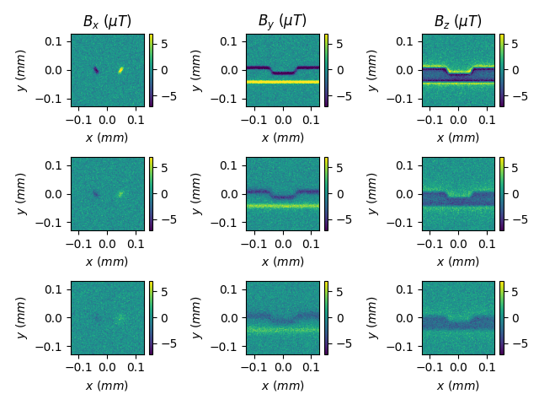}
    \caption{Simulated $B$-fields at varying measurement standoff distances; $z=1\mu m$ (top row), $z=5\mu m$ (middle row), and $z=10\mu m$ (bottom row). Each field is corrupted by $\sigma = 1.25 ]mu T$ Gaussian noise.}
    \label{fig:b-fields-simulated}
\end{figure}

From these simulated current densities we simulate $B$-field measurements at standoff distances $z = 1,5,10\mu \textrm{m}$ by applying \eqref{forward-operator}. We also add zero-mean Gaussian noise to these fields at varying strengths, ranging from $\sigma = 0.1 \mu \textrm{T}$ to $\sigma = 3 \mu \textrm{T}$. A selection of these fields is shown in Figure \ref{fig:b-fields-simulated}, where the effect of field blurring can clearly be seen with increasing standoff.  

\subsection{Implementation Details and Regularization Parameter Selection}
We compare the proposed method with a cosine tapered Fourier based reconstruction method as defined in \eqref{cos-taper}.  Applying this method requires selection of the frequency cutoff $k_{max}$. Large cutoff frequencies attempt to recover high spatial frequencies at the risk of admitting large amounts of noise into the reconstruction. Conversely, a choice of small $k_{max}$ will attenuate the noise but will also lead to a `blurry' reconstruction where high spatial frequencies are eliminated.

We choose $k_{max}$ by first defining a signal model that represents the expected decay of the Fourier spectrum of the magnetic field. Here, the current density consists of features that are $10$ microns wide, so the Fourier transform of the current distribution should decay as 
\begin{equation}
    j_{model}[k] \sim \frac{J_0}{\pi k\ 10\mu m},
\end{equation}
where $J_0$ is the nominal current density. Consequently, the Fourier spectrum of the $B$ field should decay as
\begin{equation}
    b_{model}[k] \sim \frac{J_0 g[k]}{\pi k 10\mu m},
\end{equation}
where $J_0$ is the average current density on the trace. Under the assumption that the noise is additive Gaussian white noise, the noise spectrum is flat and depends only on the variance of the added noise. For our experiments, we chose $k_{max}$ to be the frequency where the expected SNR (the ratio of the signal model spectra to the noise spectra) falls below $-6 dB$.

Solution of the regularized formulation \eqref{functional-2} was accomplished by the linearized alternating direction method of multipliers (LADMM), as implemented in \cite{Ravasi2024}. The parameter $\tau$ was chosen so that at convergence, the data residual, $||\mathcal{B}\hat{J} - \hat{\textbf{B}}||^2 \approx 3\sigma^2N^2$. This selection is motivated by the observation that if the true solution $\textbf{J}$ is recovered by the algorithm, then $3\sigma^2N^2$ is the expected value of the squared $L2$-norm residual. There are automated methods for selecting $\tau$ following this criteria \cite{Hanhela2021,Langer2017,Osadebey2014}, however, we found satisfactory results by hand tuning to the above criteria. 

\subsection{Comparison Between $L1-\textrm{curl}$ and Fourier Approaches}
\begin{figure}
    \centering
    \includegraphics[scale=.4]{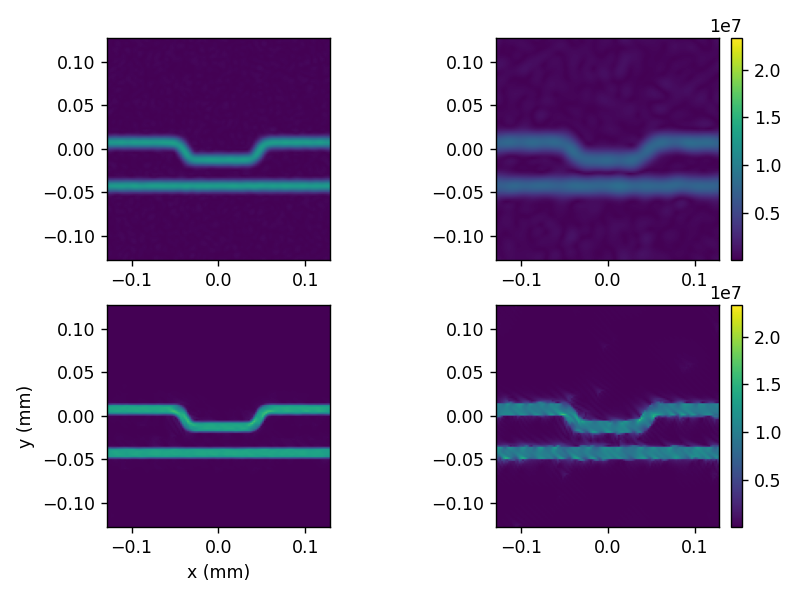}
    \caption{Recovery of simulated current distribution magnitudes $||\textbf{J}||$, $5\mu \textrm{m}$ measurement standoff. Fourier methods (top), $L1-\textrm{curl}$ method (bottom). $\sigma=0.1 \mu \textrm{T}$ (Left), $\sigma = 1.23\mu \textrm{T}$ (Right).}
    \label{fig:sim-recovery}
\end{figure}

\begin{figure}
    \centering
    \includegraphics[scale=.4]{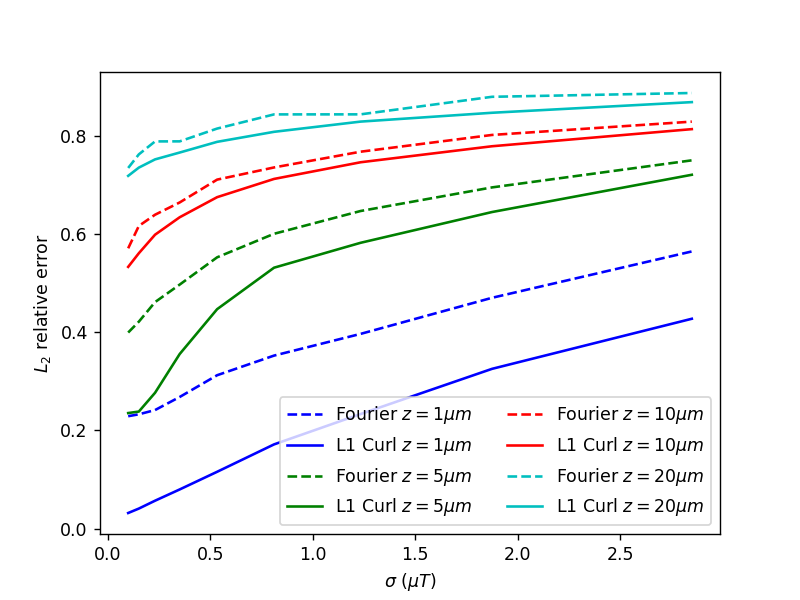}
    \caption{$L2$ relative error between recovered current density and true current density as a function of noise level and measurement standoff.}
    \label{fig:error_plots}
\end{figure}

We compared the two methods for measurement standoffs of $z=1,5,10 \mu \textrm{m}$ and noise levels ranging from $\sigma = 100 \textrm{nT}$ to $\sigma = 3 \mu \textrm{T}$. Figure \ref{fig:sim-recovery} shows a comparison between fields recovered by each method recovery for the $z = 5 \mu \textrm{m}$ measurement standoff case under two noise levels. In each case, the $L1-\textrm{curl}$ method recovers current distributions with qualitatively improved SNR and finer edge details compared to Fourier methods. Figure \ref{fig:error_plots} shows a comparison between the $L2$ relative error between the recovered current density and the true current density for each method. The $L1-\textrm{curl}$ method produces lower error across all parameters.

\subsection{Results on Experimentally Acquired Data}

\begin{figure}
    \centering
    \includegraphics[scale=1]{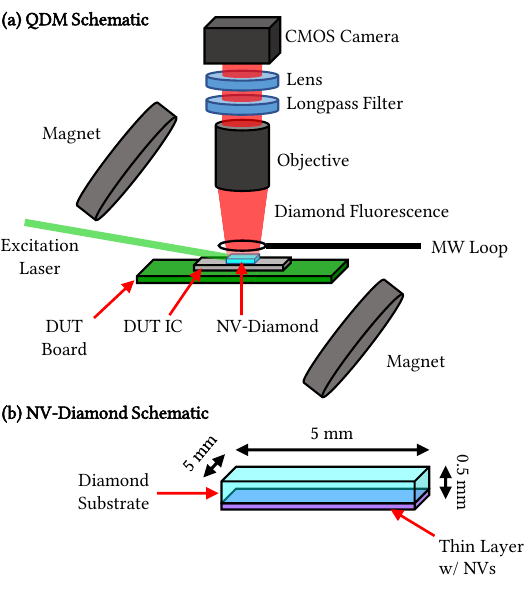}
    \caption{(a) Quantum Diamond Microscope (QDM) schematic. The instrument is built around a diamond chip
embedded with a thin layer of magnetically sensitive nitrogen-vacancy (NV) center defects which enables wide field- of- view imaging. The NV-diamond can be placed directly
on the device under test (DUT) integrated circuit (IC) to maximize spatial resolution in magnetic field images. The NV centers are excited into a magnetically sensitive state by a $532$ nm laser and a microwave (MW) loop. (b) Schematic of the NV-diamond shown in (a). These diamonds are typically $\sim 5 \textrm{mm} \times 5 \textrm{mm} \times 0.5 \textrm{mm}$ and consist of a pure diamond substrate
and a thin layer containing an ensemble of NV centers.}
    \label{fig:schematic}
\end{figure}

\begin{figure}
    \centering
    \includegraphics[scale=.5]{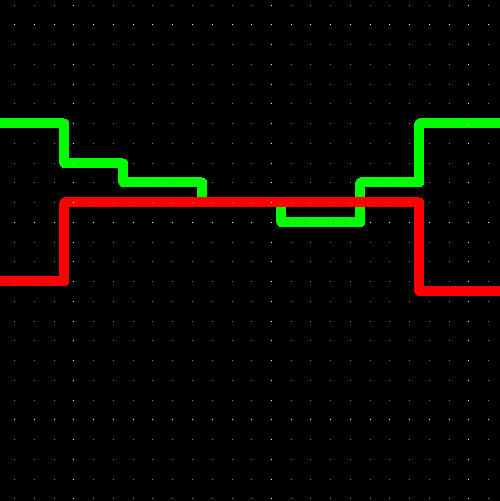}
    \caption{Schematic of PCB layout for experimental data. Top PCB layer (red) is located at approximately $65 \mu \textrm{m}$ below the measurement plane of the QDM. The bottom PCB layer (green) is located approximately $125 \mu \textrm{m}$ below the measurement plane. Each trace is approximately $0.08$ mm thick. The entire imaged region is approximately $4$ mm on each side.}
    \label{fig:pcb}
\end{figure}

\begin{figure}
    \centering
    \includegraphics[scale=.25]{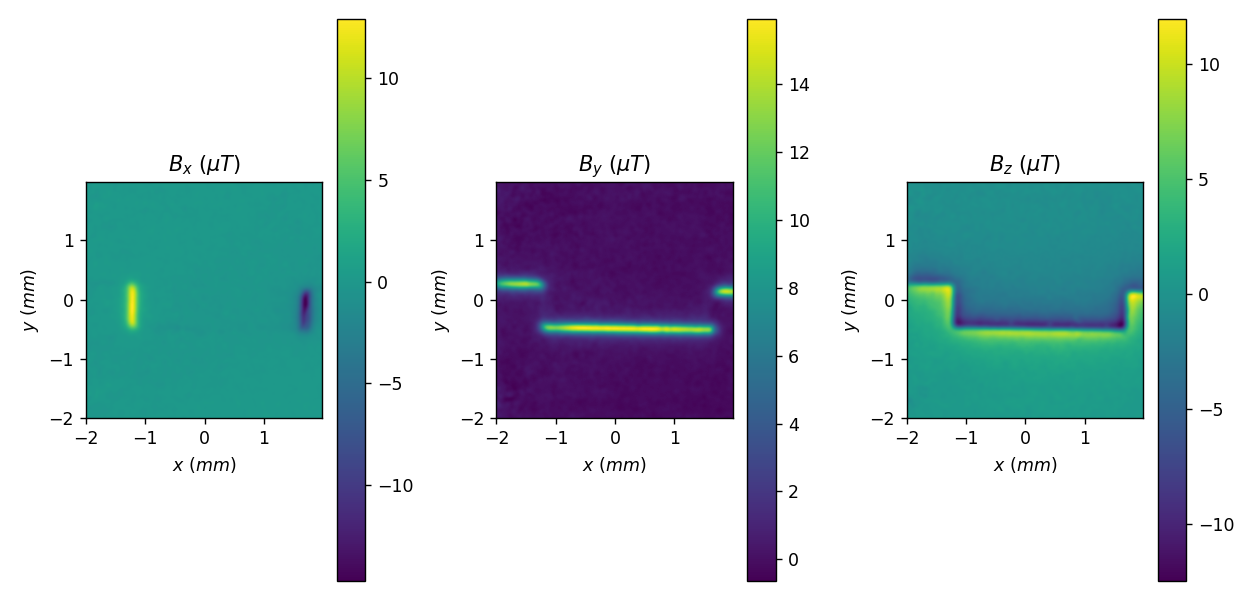}
    \includegraphics[scale=.25]{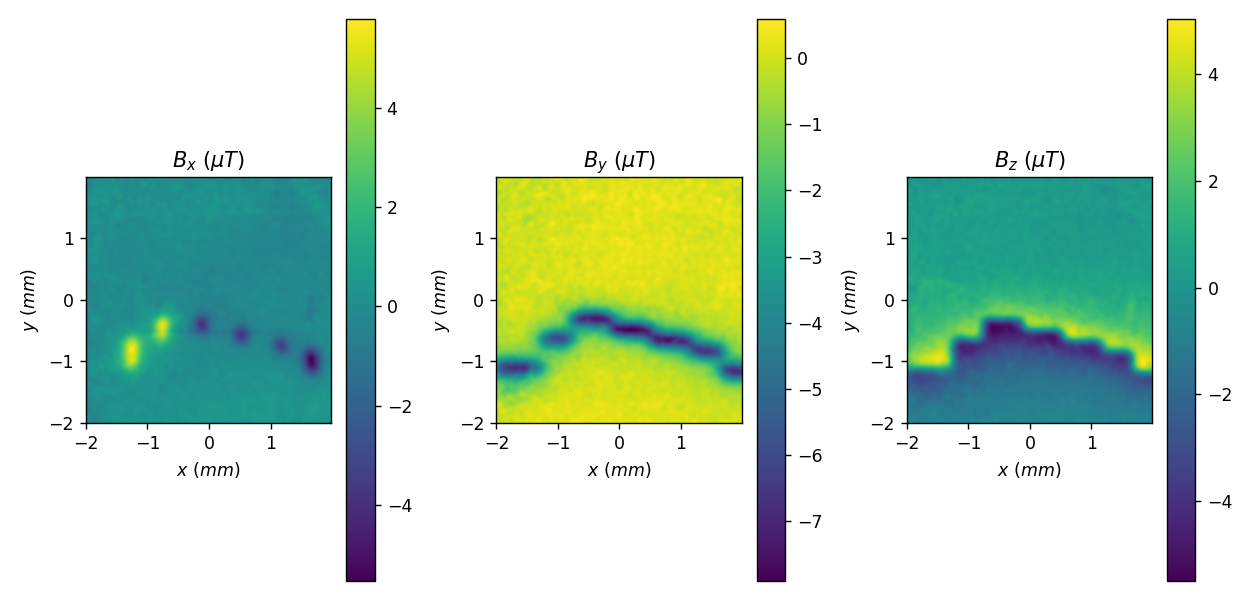}
    \caption{Measured $B$ field data from QDM placed over PCB trace region. Top layer (top). Bottom layer (bottom). Note: QDM images are rotatflipped vertically relative to the PCB layout above.}
    \label{fig:QDM-B}
\end{figure}

We also compared the $L1-\textrm{curl}$ and Fourier approaches on magnetic field data acquired by a quantum diamond microscope (QDM) magnetometer as shown in the schematic of Figure \ref{fig:schematic}. This emerging tool enables wide field-of-view, vector magnetic imaging under ambient conditions and has been used in previous studies for detecting magnetic domains in geological samples \cite{geo}, isolating failures and anomalous activity in microelectronics \cite{micro1, micro2}, and imaging cells \cite{nature-cells} among other applications. The QDM utilizes a lab-grown diamond chip embedded with an ensemble of magnetically sensitive quantum defects known as nitrogen vacancy (NV) centers. The diamond is placed directly on top of a device under test, where it is excited with a 532 nm laser and resonant microwaves. In the excitHered state, the NV centers embedded in the diamond emits a magnetic-field- dependent red fluorescence that is collected with an objective and imaged with a CMOS camera. Post processing of the spatially resolved NV center flourescence images yields vector magnetic field maps of the device under test. More detail on the specifics of the QDM can be found in \cite{levineturner} and detail on QDM vector magnetic field imaging can be found in \cite{Turner2020,oliver}. For these measurements, a diamond with a thin layer ($\sim 1 \mu m$) of NV centers ($\sim 1\mu\textrm{m}$) was laid directly onto a custom made PCB (see Figure \ref{fig:pcb}). The imaged area of the PCB contains two current carrying traces located at different depths. The top layer is located approximately $65\mu \textrm{m}$ below the measurement plane of the QDM while the bottom layer is located approximately $125 \mu \textrm{m}$ below the measurement plane. The traces were biased (one-at-a-time) with a $5 \textrm{mA}$ current and the resultant magnetic fields from the current flow wereas imaged.  Two sets of measurements were taken, one while the traces weare biased and one while they were unbiased. The vector magnetic field was computed by subtracting the unbiased from the biased data. Figure \ref{fig:QDM-B} shows the $\textbf{B}$-field data acquired by the QDM.

\begin{figure}
    \centering
    \includegraphics[scale=.4]{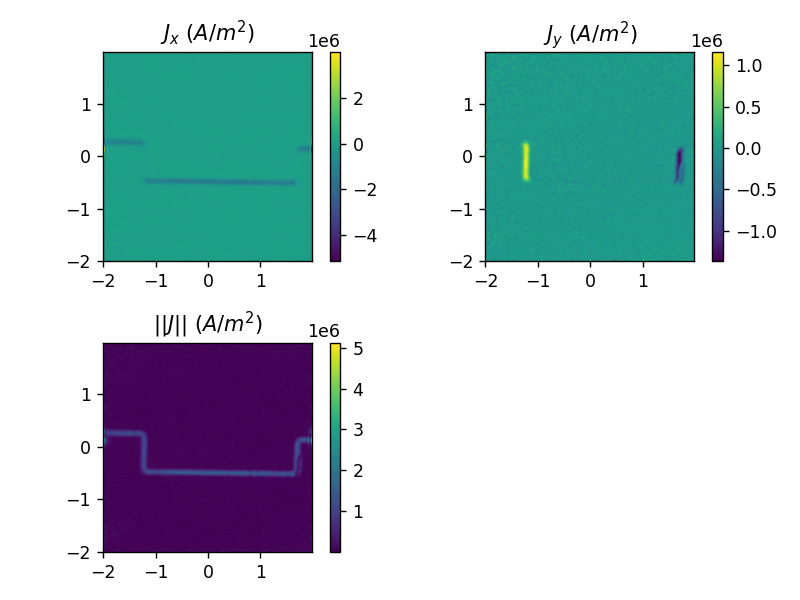}
    \includegraphics[scale=.4]{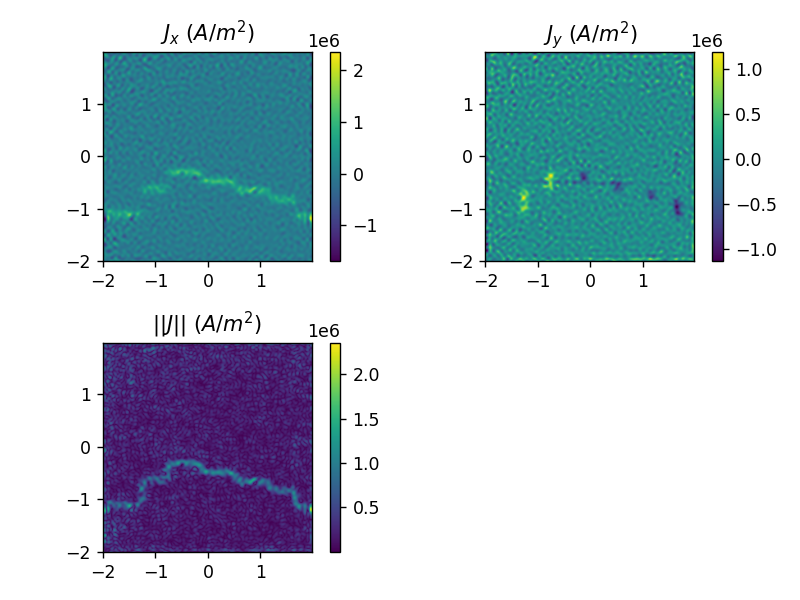}
    \caption{Recovery of current distribution by Fourier methods. Top PCB layer (top three). Bottom PCB layer (bottom three).}
    \label{fig:fourier_QDM}
\end{figure}

\begin{figure}
    \centering
    \includegraphics[scale=.4]{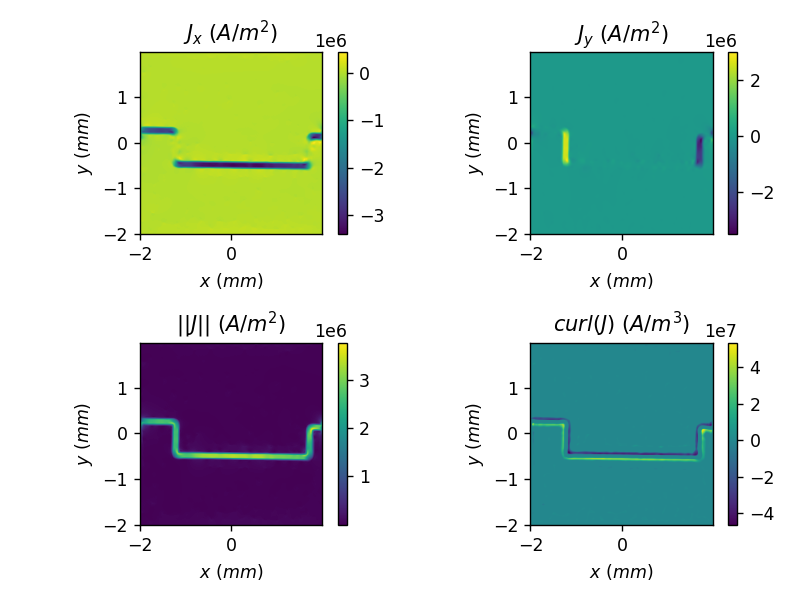}
    \includegraphics[scale=.4]{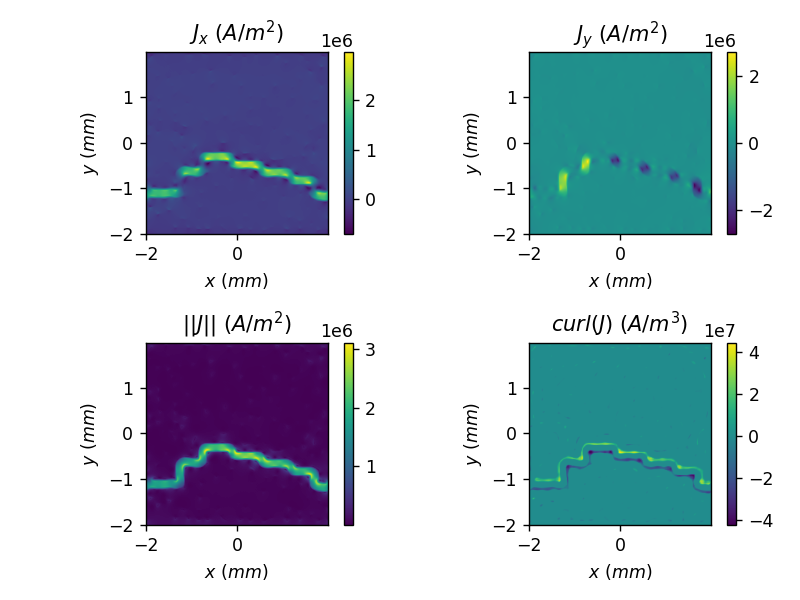}
    \caption{Recovery of current distribution by $L1-curl$ methods. Top PCB layer (top four). Bottom PCB layer (bottom four).}
    \label{fig:curl_QDM}
\end{figure}

We reconstructed current distributions from each layer using both the cosine-tapered Fourier method as well as the $L1-\textrm{curl}$ regularized method. Current density reconstructions based on the Fourier method are shown in Figure \ref{fig:fourier_QDM}. Reconstructions based on the $L1-curl$ method are shown in Figure \ref{fig:curl_QDM}. Both methods qualitatively capture the behavior of each trace; however, the $L1-\textrm{curl}$ method produced superior results as the Fourier method reconstructions contain higher noise and produce lower contrast current images. Figure \ref{fig:curl_QDM} also shows the curl of the computed current distributions, demonstrating the sparse structure of the curl.

One item we found interesting is that the phenomenon where current concentrates while passing around a corner is visible in the $L1-\textrm{curl}$ based reconstructions. This is something that should be observed since current will prefer to take a shorter path through a homogeneous conductor. This effect is not visible in the Fourier based reconstructions owing to the sacrificing of high-spatial frequency information. The capability of imaging this phenomenon is similar to the results published in \cite{Ku2020Nature}. We beleive that our technique may allow for imaging of current hydrodynamic phenomenologies using less sensitive instruments and longer measurement standoff than was previously possible. 

\section{Conclusion and Future Work}

We presented a method for recovery of two-dimensional current distributions from magnetic field data. This method represents a substantial improvement over Fourier-based methods. We believe that this method has the potential to improve microelectronic inspection procedures based on magnetic current imaging. In particular we plan on investigating extension of the proposed approach to magnetic current imaging of $3-d$ stacked circuit architectures. 

We also believe that this method will allow for imaging of hydrodynamic current flow which will allow for may in turn pave the way to study of hydrodynamic transport in quantum critical fluids with potential application to the study of high-Tc superconductors \cite{Ku2020Nature}. In subsequent work, we plan to investigate the applicability of this technique to multi-layer current distributions representative of those found in more complicated microelectronic devices.

\section{Acknowledgements}
The view, opinions, and/or findings in this report are those of The MITRE Corporation and should not be interpreted as
representing the official views or policies of the Department of Defense or the U.S. Government. This technical data was developed using contract funds under Basic Contract No. W56KGU-22-F-0017.
Approved for Public Release; Distribution Unlimited. Public Release Case Number 24-1623.
\textcopyright~2024 The MITRE Corporation.

\appendix
\renewcommand{\thesection}{\Alph{section}}
\numberwithin{equation}{section}
\section{Fourier Transform Conventions}\label{app:A}
The Fourier transform of a function $F\in L^1(\mathbb{R})$ is defined by
\begin{equation}
f(\xi) = \int_{-\infty}^\infty F(x) \exp(-2\pi i x \xi) dx,
\end{equation}
and the corresponding transform of a function $F(x,y)\in L^1(\mathbb{R}^2)$ is
\begin{equation}
f(\textbf{k} = [k_x,k_y]^t) = \int\int_{\mathbb{R}^2} F(x,y) \exp(-2\pi i (xk_x + yk_y)) dx dy.
\end{equation}
If a function $F(x)$ has support only on the interval $[-M/2,M/2]$ and $F$ is sampled at discrete locations $F[n] \equiv F(nT)$ then the discrete Fourier transform is given by
\begin{equation}
f[k_x] = \sum_{n}F[n]\exp(-2\pi i (nk_x)/N),
\end{equation}
where we have made the simplifying assumption that $T$ divides $M$ evenly and $N = M/T$. By the Poisson summation formula,
\begin{gather}
\begin{split}
    f[k_x] = T^{-1} f_{1/T}(k_x/M),
    \\
    f_{1/T}(k_x) = \sum_{k=-\infty}^\infty f(k_x + k/T).
\end{split}
\end{gather}
\section{Fourier Transform of the Magnetic Convolution Kernel}\label{app:B}
Define
\begin{equation}
    \hat{G}(\textbf{x},z) = \frac{\mu_0d}{4\pi}\frac{1}{(||\textbf{x}||^2 + z^2)^{3/2}}.
\end{equation}
The Fourier transform of $\hat{G}$ can be derived by noting that the kernel $\hat{G}$ is radially symmetric and converting the Fourier transform into a polar form as follows,
\begin{gather}
    \begin{split}
        \hat{g}(\textbf{k},z) = \int_{\mathbb{R^2}}\hat{G}(\textbf{x},z)\exp(-2\pi i(xk_x + yk_y))\ dx\ dy
        \\
        = \int_0^\infty \hat{G}(r',z)\int_0^{2\pi}\exp(-2\pi i r' ||\textbf{k}||(\cos\theta\cos k_\theta - \sin\theta\sin k_\theta))r'\ dr'\ d\theta
        \\
        = \int_0^\infty \hat{G}(r',z)\int_0^{2\pi}\exp(-2\pi i r'||\textbf{k}||\cos(k_\theta - \theta))r'\ dr\ d\theta
        \\
        = \int_0^\infty \hat{G}(r',z)\int_0^{2\pi}\exp(-2\pi i r'||\textbf{k}||\cos(\theta))r'\ dr\ d\theta
        \\
        =2\pi\int_0^\infty \hat{G}(r',z)J_0(2\pi ||\textbf{k}|| r')r'\ dr'
        \\
        =2\pi\ \left(\frac{\mu_0d}{4\pi}\right)\int_0^{\infty}\frac{1}{(r'^2 + z^2)^{3/2}}J_0(2\pi ||\textbf{k}||r')r'\ dr'
        \\
        = \frac{\mu_0d}{2}z^{-1}\exp(-2\pi z ||\textbf{k}||).
    \end{split}
\end{gather}
For the last equality we employ Equation (4) of section 6.554 from \cite{table-of-integrals}. 

We also compute the Fourier transform of $x\hat{G}$ using the relation $\mathcal{F}(xF(x))(k) = \frac{i}{2\pi}\frac{d}{dk}\mathcal{F}(f)(k)$. Thus,
\begin{equation}\label{ft-gx}
    \mathcal{F}(x\hat{G}) = \frac{-i\mu_0d}{2}\exp(-2\pi z||\textbf{k}||)\frac{k_x}{||k||}.
\end{equation}
\bibliography{references}

\begin{thebibliography}{10}
\expandafter\ifx\csname url\endcsname\relax
  \def\url#1{\texttt{#1}}\fi
\expandafter\ifx\csname urlprefix\endcsname\relax\def\urlprefix{URL }\fi
\expandafter\ifx\csname href\endcsname\relax
  \def\href#1#2{#2} \def\path#1{#1}\fi

\bibitem{rudin}
L.~I. Rudin, S.~Osher, E.~Fatemi, Nonlinear total variation based noise removal
  algorithms, Physica D: Nonlinear Phenomena 60~(1) (1992) 259--268.
\newblock \href {https://doi.org/10.1016/0167-2789(92)90242-F}
  {\path{doi:10.1016/0167-2789(92)90242-F}}.

\bibitem{chambolle}
A.~Chambolle, An algorithm for total variation minimization and applications,
  Journal of Mathematical Imaging and Vision 20 (2004) 89--97.

\bibitem{donoho}
D.~L. Donoho, For most large underdetermined systems of linear equations the
  minimal l1‐norm solution is also the sparsest solution, Communications on
  Pure and Applied Mathematics 59 (2006).

\bibitem{tao}
E.~Candes, J.~Romberg, T.~Tao, Stable signal recovery from incomplete and
  inaccurate measurements (2005).
\newblock \href {http://arxiv.org/abs/math/0503066}
  {\path{arXiv:math/0503066}}.

\bibitem{lemarie}
Lemarie-Rieusset, Analyses multi-resolutions non orthogonales, commutation
  entre projecteurs et derivation et ondelettes vecteurs a divergence nulle,
  Revista Matem\'{a}tica Iberoamericana 8~(2) (1992) 221 -- 237.

\bibitem{Harouna2013}
S.~K. Harouna, V.~Perrier, S.~K. Harouna, V.~Perrier, Effective construction of
  divergence-free wavelets on the square, Journal of Computational and Applied
  Mathematics 240 (2013) 74--86.
\newblock \href {https://doi.org/10.1016/j.cam.2012.07.029ï}
  {\path{doi:10.1016/j.cam.2012.07.029ï}}.

\bibitem{kadri}
S.~K. Harouna, P.~Dérian, P.~Héas, E.~Mémin, S.~Kadri-Harouna, P.~Dérian,
  P.~Héas, E.~Mémin, Divergence-free wavelets and high order regularization,
  International Journal of Computer Vision 103 (2013) 80--99.
\newblock \href {https://doi.org/10.1007/s11263-012-0595-7ï}
  {\path{doi:10.1007/s11263-012-0595-7ï}}.

\bibitem{Turner2020}
M.~J. Turner, N.~Langellier, R.~Bainbridge, D.~Walters, S.~Meesala, T.~M.
  Babinec, P.~Kehayias, A.~Yacoby, E.~Hu, M.~Lon{\v{c}}ar, R.~L. Walsworth,
  E.~V. Levine, {Magnetic Field Fingerprinting of Integrated-Circuit Activity
  with a Quantum Diamond Microscope}, Physical Review Applied 14~(1) (2020)
  1--12.
\newblock \href {http://arxiv.org/abs/2004.03707} {\path{arXiv:2004.03707}},
  \href {https://doi.org/10.1103/PhysRevApplied.14.014097}
  {\path{doi:10.1103/PhysRevApplied.14.014097}}.

\bibitem{levineturner}
E.~V. Levine, M.~J. Turner, P.~Kehayias, C.~A. Hart, N.~Langellier, R.~Trubko,
  D.~R. Glenn, R.~R. Fu, R.~L. Walsworth, Principles and techniques of the
  quantum diamond microscope, Nanophotonics 8~(11) (2019) 1945--1973.
\newblock \href {https://doi.org/doi:10.1515/nanoph-2019-0209}
  {\path{doi:doi:10.1515/nanoph-2019-0209}}.

\bibitem{wellstood}
F.~C. Wellstood, J.~Matthews, S.~Chatraphorn, Ultimate limits to magnetic
  imaging, IEEE Transactions on Applied Superconductivity 13~(2) (2003)
  258--260.
\newblock \href {https://doi.org/10.1109/TASC.2003.813699}
  {\path{doi:10.1109/TASC.2003.813699}}.

\bibitem{roth}
B.~J. Roth, N.~G. Sepulveda, J.~P. Wikswo, Using a magnetometer to image a
  two‐dimensional current distribution, Journal of Applied Physics 65~(1)
  (1989) 361--372.
\newblock \href {https://doi.org/10.1063/1.342549}
  {\path{doi:10.1063/1.342549}}.

\bibitem{Heitz2010VariationalFF}
D.~Heitz, {\'E}.~M{\'e}min, C.~Schn{\"o}rr,
  \href{https://api.semanticscholar.org/CorpusID:17236621}{Variational fluid
  flow measurements from image sequences: synopsis and perspectives},
  Experiments in Fluids 48 (2010) 369--393.
\newline\urlprefix\url{https://api.semanticscholar.org/CorpusID:17236621}

\bibitem{mallat2008wavelet}
S.~Mallat, A Wavelet Tour of Signal Processing: The Sparse Way, Elsevier
  Science, 2008.

\bibitem{fipy}
J.~E. Guyer, D.~Wheeler, J.~A. Warren, Fipy: Partial differential equations
  with python, Computing in Science \& Engineering 11~(3) (2009) 6--15.
\newblock \href {https://doi.org/10.1109/MCSE.2009.52}
  {\path{doi:10.1109/MCSE.2009.52}}.

\bibitem{Ravasi2024}
M.~Ravasi, M.~V. Örnhag, N.~Luiken, O.~Leblanc, E.~Uruñuela, Pyproximal -
  scalable convex optimization in python, Journal of Open Source Software
  9~(95) (2024) 6326.
\newblock \href {https://doi.org/10.21105/joss.06326}
  {\path{doi:10.21105/joss.06326}}.

\bibitem{Hanhela2021}
M.~Hanhela, O.~Gröhn, M.~Kettunen, K.~Niinimäki, M.~Vauhkonen,
  V.~Kolehmainen, Data-driven regularization parameter selection in dynamic
  mri, Journal of Imaging 7 (2 2021).
\newblock \href {https://doi.org/10.3390/jimaging7020038}
  {\path{doi:10.3390/jimaging7020038}}.

\bibitem{Langer2017}
A.~Langer, Automated parameter selection for total variation minimization in
  image restoration, Journal of Mathematical Imaging and Vision 57 (2017)
  239--268.
\newblock \href {https://doi.org/10.1007/s10851-016-0676-2}
  {\path{doi:10.1007/s10851-016-0676-2}}.

\bibitem{Osadebey2014}
M.~Osadebey, N.~Bouguila, D.~Arnold, Optimal selection of regularization
  parameter in total variation method for reducing noise in magnetic resonance
  images of the brain, Biomedical Engineering Letters 4 (2014) 80--92.
\newblock \href {https://doi.org/10.1007/s13534-014-0126-2}
  {\path{doi:10.1007/s13534-014-0126-2}}.

\bibitem{geo}
D.~R. Glenn, R.~R. Fu, P.~Kehayias, D.~Le~Sage, E.~A. Lima, B.~P. Weiss, R.~L.
  Walsworth, Micrometer-scale magnetic imaging of geological samples using a
  quantum diamond microscope, Geochemistry, Geophysics, Geosystems 18~(8)
  (2017) 3254--3267.
\newblock \href {https://doi.org/10.1002/2017GC006946}
  {\path{doi:10.1002/2017GC006946}}.

\bibitem{micro1}
M.~Ashok, M.~J. Turner, R.~L. Walsworth, E.~V. Levine, A.~P. Chandrakasan,
  Hardware trojan detection using unsupervised deep learning on quantum diamond
  microscope magnetic field images, J. Emerg. Technol. Comput. Syst. 18~(4)
  (oct 2022).
\newblock \href {https://doi.org/10.1145/3531010} {\path{doi:10.1145/3531010}}.

\bibitem{micro2}
P.~Kehayias, M.~A. Delaney, R.~A. Haltli, S.~M. Clark, M.~C. Revelle, A.~M.
  Mounce, Fault localization in a microfabricated surface ion trap using
  diamond nitrogen-vacancy center magnetometry (2024).
\newblock \href {http://arxiv.org/abs/2403.08731} {\path{arXiv:2403.08731}}.

\bibitem{nature-cells}
D.~R. Glenn, K.~Lee, H.~Park, R.~Weissleder, A.~Yacoby, M.~D. Lukin, H.~Lee,
  R.~L. Walsworth, C.~B. Connolly, Single-cell magnetic imaging using a quantum
  diamond microscope, Nature Methods 12~(8) (2015) 736--738.
\newblock \href {https://doi.org/10.1038/nmeth.3449}
  {\path{doi:10.1038/nmeth.3449}}.

\bibitem{oliver}
S.~M. Oliver, D.~J. Martynowych, M.~J. Turner, D.~A. Hopper, R.~L. Walsworth,
  E.~V. Levine, {Vector Magnetic Current Imaging of an 8 nm Process Node Chip
  and 3D Current Distributions Using the Quantum Diamond Microscope}, Vol.
  ISTFA 2021: Conference Proceedings from the 47th International Symposium for
  Testing and Failure Analysis of International Symposium for Testing and
  Failure Analysis, 2021, pp. 96--107.
\newblock \href {https://doi.org/10.31399/asm.cp.istfa2021p0096}
  {\path{doi:10.31399/asm.cp.istfa2021p0096}}.

\bibitem{Ku2020Nature}
M.~J.~H. Ku, T.~X. Zhou, Q.~Li, Y.~J. Shin, J.~K. Shi, C.~Burch, L.~E.
  Anderson, A.~T. Pierce, Y.~Xie, A.~Hamo, U.~Vool, H.~Zhang, F.~Casola,
  T.~Taniguchi, K.~Watanabe, M.~M. Fogler, P.~Kim, A.~Yacoby, R.~L. Walsworth,
  Imaging viscous flow of the dirac fluid in graphene, Nature 583~(7817) (2020)
  537--541.
\newblock \href {https://doi.org/10.1038/s41586-020-2507-2}
  {\path{doi:10.1038/s41586-020-2507-2}}.

\bibitem{table-of-integrals}
I.~Gradshteyn, I.~Ryzhik, Table of Integrals, Series, and Products, 7th
  Edition, Academic Press, 2007.

\end{thebibliography}

\end{document}